# TIME SERIES ANALYSIS METHODS APPLIED TO THE SUPER-KAMIOKANDE I DATA

Gioacchino Ranucci
Istituto Nazionale di Fisica Nucleare
Via Celoria 16 - 20133 Milano
e-mail: gioacchino.ranucci@mi.infn.it

ABSTRACT

The need to unravel modulations hidden in noisy time series of experimental data is a well known problem, traditionally attacked through a variety of methods, among which a popular tool is the so called Lomb-Scargle periodogram. Recently, for a class of problems in the solar neutrino field, it has been proposed an alternative maximum likelihood based approach, intended to overcome some intrinsic limitations affecting the Lomb-Scargle implementation. This work is focused to highlight the features of the likelihood methodology, introducing in particular an analytical approach to assess the quantitative significance of the potential modulation signals. As an example, the proposed method is applied to the time series of the measured values of the $^{8}B$ neutrino flux released by the Super-Kamiokande collaboration, and the results compared with those of previous analysis performed on the same data sets. It is also examined in detail the comparison between the Lomb-Scargle and the likelihood methods, giving in the appendix the complete demonstration of their close relationship.

## 1.Introduction

The Lomb-Scargle periodogram method [1][2] is a very popular tool to scan experimental data organised in time series in order to highlight the presence of periodicities masked by overwhelming noise. Since in its classical implementation it does not make use of the error estimates associated with each data point in the time series, it has been suggested [3] the possibility to adopt an alternative likelihood approach to exploit fully the error information.

On the other hand, the Lomb-Scargle methodology features the interesting property that a simple "false alarm" probability formula can be invoked to assess the significance of the modulation signature found in the estimated spectrum, whereas in the case of application of the likelihood test the same significance assessment has been done through Monte Carlo evaluations.

The main focus of this work, besides reviewing the overall characteristics of the likelihood methodology in the framework of time series scan, is to show how also for the likelihood test a "false alarm" probability formula can be computed and used to evaluate the significance of the potential modulation features identified in the spectrum, in total analogy with the Lomb-Scargle method, thus avoiding the need to invoke the use of the Monte Carlo significance assessment.

In the second part of this work this methodology is then extensively applied to the scan of both the 10 and 5 day binned time series of the $^8$B solar neutrino data published by the Super-Kamiokande collaboration, which are well suited to undergo such an analysis method.

Finally, in the appendix it is shown how to modify the Lomb-Scargle periodogram in order to fully exploit also in the framework of this method the errors associated with each data point, and how the resulting modified periodogram compares with the likelihood based spectrum.

## 2. Some remarks on the application of the likelihood ratio test to disentangle time modulations in white noise

The problem of detecting whether or not a signal is present in overwhelming noise is ubiquitous in science. Several detection strategies, thoroughly described in the literature, lead to a likelihood ratio test against a threshold, whose value is defined according to which hypothesis testing criterion is chosen [4].

The likelihood ratio is evaluated dividing each other the likelihood functions related to the two different situations to be confronted, i.e. presence of the signal against absence of the signal. In many signal detection applications this is done knowing a priori which are the characteristics of the signal which could be embedded in the noise, as well as the features of the noise itself. The decision system is thus engineered starting from these building blocks; in practice this means that a likelihood ratio is written which incorporates all the prior knowledge on the random process under investigation: the signal is then declared present if the ratio is greater than the pre-defined threshold, or absent if the ratio is less than it.

The performances of such an inference procedure are quantified by two factors of merit: the probability of false alarm (the quantity we are interested to), which gives the fractions of times that the decision system is confused by noise fluctuations mimicking the signal, and the detection efficiency, which expresses the fractions of times that the system correctly identifies a signal actually present.

In the specific application of this work, a simple criterion particularly suited to produce an effective and useful expression for the false alarm probability is that of the so called ideal observer, in which the threshold for the likelihood ratio is set equal to 1: hence, for a given experimental outcome, the inference about the presence either of signal plus noise or of noise alone is taken on the basis of which of the two corresponding likelihood functions has the highest value.

In order to be specific to the problem we are dealing with, let's assume that the process is sampled at $k$ time points, and denote with $x_k$ the samples, with $X_k$ the values assumed at the sampling points by the hypothetical signal whose presence in the process has to be tested, and with $\sigma_k$ the noise variance at each sampling point. In case of white gaussian noise with zero mean value, the likelihood ratio is then

$$L = \frac{e^{-\frac{1}{2}\left(\sum\limits_{k=1}^{N}\frac{(x_k - X_k)^2}{\sigma_k^2}\right)}}{e^{-\frac{1}{2}\left(\sum\limits_{k=1}^{N}\frac{x_k^2}{\sigma_k^2}\right)}} \qquad (1).$$

As usual the log-likelihood ratio is more convenient for the analysis; we can thus write

$$S = \log L = \frac{1}{2}\left(\sum_{k=1}^{N}\frac{x_k^2}{\sigma_k^2}\right) - \frac{1}{2}\left(\sum_{k=1}^{N}\frac{(x_k - X_k)^2}{\sigma_k^2}\right) \qquad (2).$$

Hence, in a system where the functional form of the searched signal is known, the log-likelihood ratio test reduces to evaluate the expression (2), opting for the presence of the signal if it is greater than 0, or for its absence if the same expression amounts to a negative value.

We pass now to the calculation of the false alarm probability, i.e. the probability to claim the presence of the signal while it is actually absent.

To show in a simple way how this probability can be computed, let's write the (2) as comparison to zero and perform some subsequent manipulations

$$\frac{1}{2}\left(\sum_{k=1}^{N}\frac{x_k^2}{\sigma_k^2}\right) - \frac{1}{2}\left(\sum_{k=1}^{N}\frac{(x_k - X_k)^2}{\sigma_k^2}\right) \geq 0 \qquad (3)$$

$$\sum_{k=1}^{N}\left(\frac{x_k^2}{2\sigma_k^2} - \frac{x_k^2}{2\sigma_k^2} + \frac{xX_k}{\sigma_k^2} - \frac{X_k^2}{2\sigma_k^2}\right) \geq 0 \qquad (4)$$

$$\sum_{k=1}^{N}\left(\frac{x_k X_k}{\sigma_k^2} - \frac{X_k^2}{2\sigma_k^2}\right) \geq 0 \qquad (5)$$

This equation can be written in an expressive way by isolating the random variables from the fixed numbers at the two different sides of the inequality (5)

$$\sum_{k=1}^{N}\frac{x_k X_k}{\sigma_k^2} \geq \sum_{K=1}^{N}\frac{X_k^2}{2\sigma_k^2} \qquad (6)$$

The inequality written in such a way states that, for a given signal shape represented by its samples $X_k$, and for a given noise characteristic, described by the noise variance at the sampling points, the likelihood ratio test is equivalent to compare the random quantity on the left side with the threshold on the right side of (6), i.e.

$$th = \frac{1}{2}\sum_{K=1}^{N}\frac{X_k^2}{\sigma_k^2} \text{ or } th = \frac{1}{2}\sigma_{tot}^2 \text{ with } \sigma_{tot}^2 = \sum_{K=1}^{N}\frac{X_k^2}{\sigma_k^2} \quad (7).$$

The reason why the above sum is denoted with $\sigma_{tot}^2$ will become clear in the following.

In order to evaluate the false alarm probability we have to consider the occurrences in which the samples $x_k$ are actually only composed by noise, but with the combination expressed by the left side of (6) overcoming the threshold *th* because of large fluctuations. The inspection of the (6) shows that in such a situation the random quantity on the left side is constituted by the weighted sum of the gaussian, with zero mean value, random variables $x_k$, each characterized by the variance $\sigma_k^2$. Simple probabilistic rules tell that the overall sum is distributed, as well, according to a zero centered normal distribution, whose variance is the weighted sum of the individual variances, with weights equal to the square of the weights in (6). Thus we have

$$\sigma_{tot}^2 = \sum_{k=1}^{N}\left(\frac{X_k}{\sigma_k^2}\right)^2 \sigma_k^2 \quad (8)$$

and finally

$$\sigma_{tot}^2 = \sum_{k=1}^{N}\frac{X_k^2}{\sigma_k^2} \quad (9).$$

Therefore, the false alarm probability is simply given by the integral above the threshold $th=0.5\ \sigma^2_{tot}$ of the normal distribution with zero mean value and variance $\sigma^2_{tot}$; hence we have

$$P_{fa} = \int_{th}^{\infty}\frac{1}{\sqrt{2\pi\sigma_{tot}^2}}e^{-\frac{t^2}{2\sigma_{tot}^2}}dt \quad (10).$$

It can be easily shown that this integral can be conveniently written in a very concise form as

$$P_{fa} = \frac{1}{2}erfc\left(\frac{\sigma_{tot}}{2\sqrt{2}}\right) \quad (11)$$

where *erfc* is the complemented error function.

**3. Specialization to the case of sinusoidal signals oscillating around a baseline**

The results got in the previous paragraph are very general, in the sense that there is actually no theoretical limitation on the shape of the signal to be searched for, which is defined in a very general way simply through the sequence of the $X_k$ values.

Since in many cases time series data are investigated for the presence of periodic sinusoidal signals, it is thus useful to specialize the methodology of paragraph 2 for the search of a sinusoid, buried in white noise, fluctuating around a baseline (in the case dealt with in the next paragraphs it will be clear that the baseline represents the average neutrino flux).

Let's denote this baseline with $F$ and the signal oscillating around it, evaluated at the sampling points, as

$$S_k = AF\sin(2\pi f t_{w,k} + \varphi) \tag{12}$$

$A$ being thus the amplitude of the oscillation, expressed as fraction of the average flux $F$; $\varphi$ is the phase, and the $t_k$ are the times at which the process is sampled.

The (2), then, is transformed into

$$S = \log L = \frac{1}{2}\left(\sum_{k=1}^{N}\frac{(x_k - F)^2}{\sigma_k^2}\right) - \frac{1}{2}\left(\sum_{k=1}^{N}\frac{(x_k - F - S_k)^2}{\sigma_k^2}\right) \qquad (13).$$

Starting from this last expression, it can be shown that the (6) now becomes

$$\sum_{k=1}^{N}\frac{(x_k - F)S_k}{\sigma_k^2} \geq \sum_{K=1}^{N}\frac{S_k^2}{2\sigma_k^2} \tag{14}$$

where, in case of pure noise with no periodic signal in it, the variables $x_k$ - $F$ are gaussian distributed with zero mean value. Following the arguments of the previous paragraph, we have, thus, for the false alarm probability the same formula (11), in which the quantity $\sigma_{tot}$ is now given by

$$\sigma_{tot}^2 = \sum_{k=1}^{N}\frac{S_k^2}{\sigma_k^2} \qquad (15).$$

It is worth to remind that the derivations of the present and previous paragraph are based on the assumption of known signal shapes, which in particular means that the parameters $F$, $\varphi$, $A$ and $f$ are all to be considered as known quantities.

Let’s now turn to our original inverse problem, in which we deal with a time variation of a priori unknown frequency, amplitude and phase, around a baseline $F$ that, on the other hand, is supposed known. In this framework, the likelihood ratio test can be specialized to a modulation search algorithm through a two steps implementation: first, a scan is performed over a frequency window in which the modulation signature is searched for, and for each investigated frequency the best sinusoidal fit to the data is determined via the maximization of the respective likelihood function (in practice, the numerator of the likelihood ratio or the second factor of the log-likelihood ratio) by proper choices of the unknown amplitude and phase parameters characterizing the periodic time signal; afterwards, to the identified potential modulations it is attributed a significance expressed by the (11), in which $\sigma_{tot}$ is given by the (15), or, according to more conventional standards, by the quantity

$$C.L.(\%) = (1 - P_{fa}) \times 100 \tag{16}$$

that can be interpreted as the confidence level for the presence of a true modulation at each scanned frequency.

The spectrum $S(f)$ of the random process is conventionally assumed as the log-likelihood ratio (13) evaluated at each frequency by inserting the corresponding amplitude and phase parameters which maximized the likelihood function.

In the following it will be shown the outcome of the application of this methodology to the Super-Kamiokande I 5 and 10 day binned data.

## 4. Application to the analysis of the 10 day binned Super-Kamiokande I data

The Super-Kamiokande collaboration published the time series of the $^8$B neutrino flux measurements organized both in 10 and 5 day bins [5]. For each bin it was provided, together with the flux value, also the respective mean live $t_{wk}$, properly evaluated in order to take into account the live time of the detector, as well as the relevant corrective factor to remove the 7% peak-to-peak annual variation due to the Earth's orbit eccentricity around the Sun. All the calculations reported in the following have been performed after the application of such a correction.

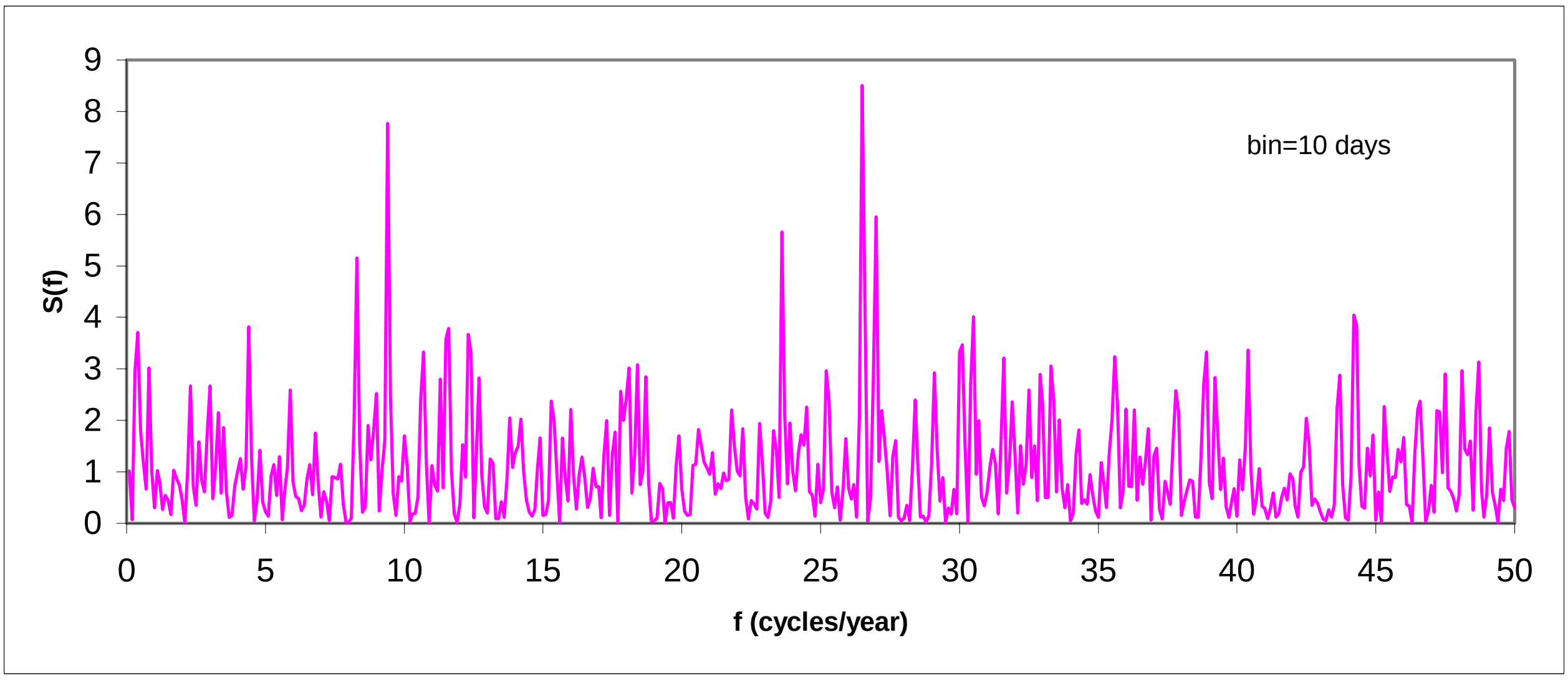

**Figure 1** - *Spectrum of the 10 day binned Super-Kamiokande data series*

Since the Super-Kamiokande collaboration published for each bin the two asymmetric errors, then the $\sigma_k$ are taken as

$$\sigma_k = \frac{\sigma_k^{up} + \sigma_k^{down}}{2} \qquad (17).$$

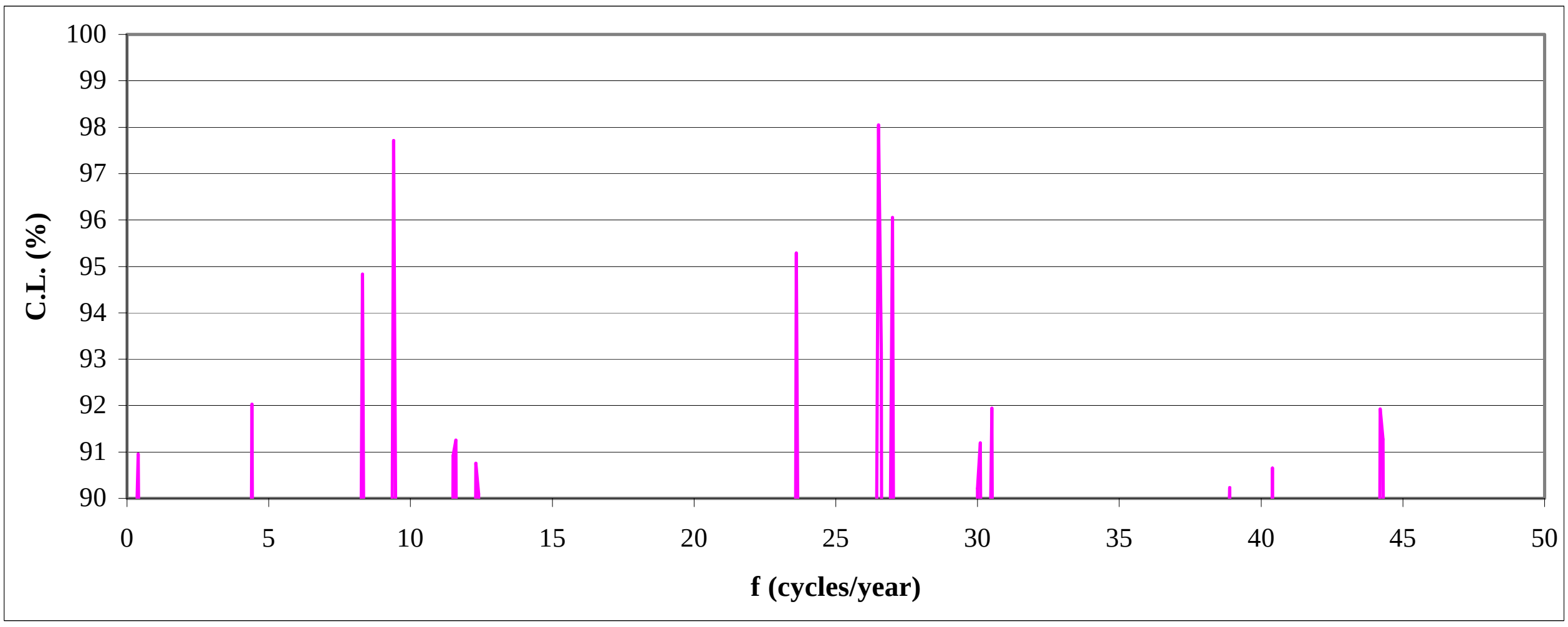

**Figure 2** - *Confidence levels of the more prominent lines in the 10 day spectrum*

It should be outlined that the value $F$ of the average neutrino flux to be inserted in the (13) is obtained by averaging the measured values at all bins.

In Fig. 1 the plot of the spectrum *S(f)*, obtained as explained at the end of the previous paragraph, is reported for the 10 day binned data, plotted over the frequency range spanning from 0 to 50 cycles/year, while in Fig. 2 the confidence levels for the more pronounced line features identified in the spectrum are shown, as deduced from the (16).

From Fig. 2 we see that actually only 4 lines can be considered of interesting statistical significance, overcoming the 95% confidence level. Particular attention is obviously drawn from the two highest confidence level lines, located respectively at 9.4 cycles/year (97.7% C.L.) and 26.52 cycles/year (98.1% C.L.) (note: here and in the following to each line should be considered attached an uncertainty of ± 0.15 cycles/year, evaluated from the FWHM of the spectral lines).

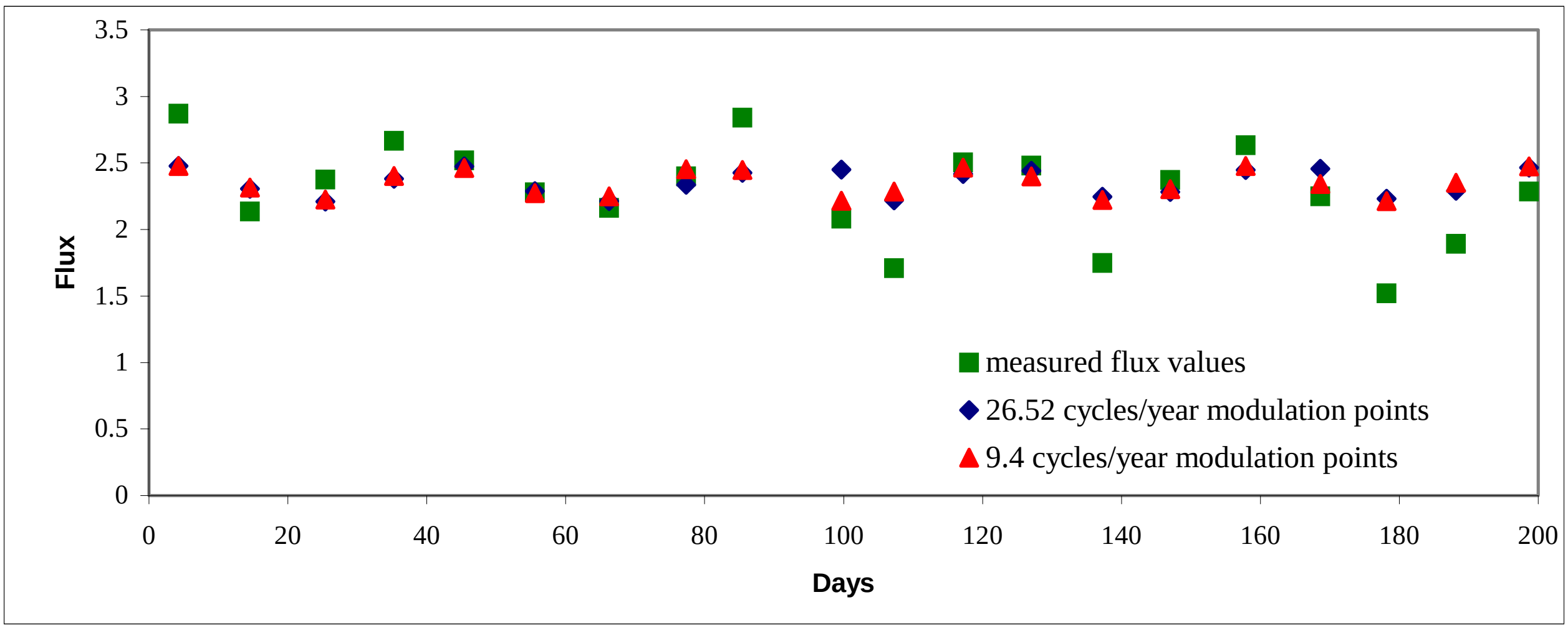

**Figure 3** - *The alias phenomenon manifests with the high degree of overlap of the points of the 26.52 and 9.4 modulations evaluated at the sampling times of the series*

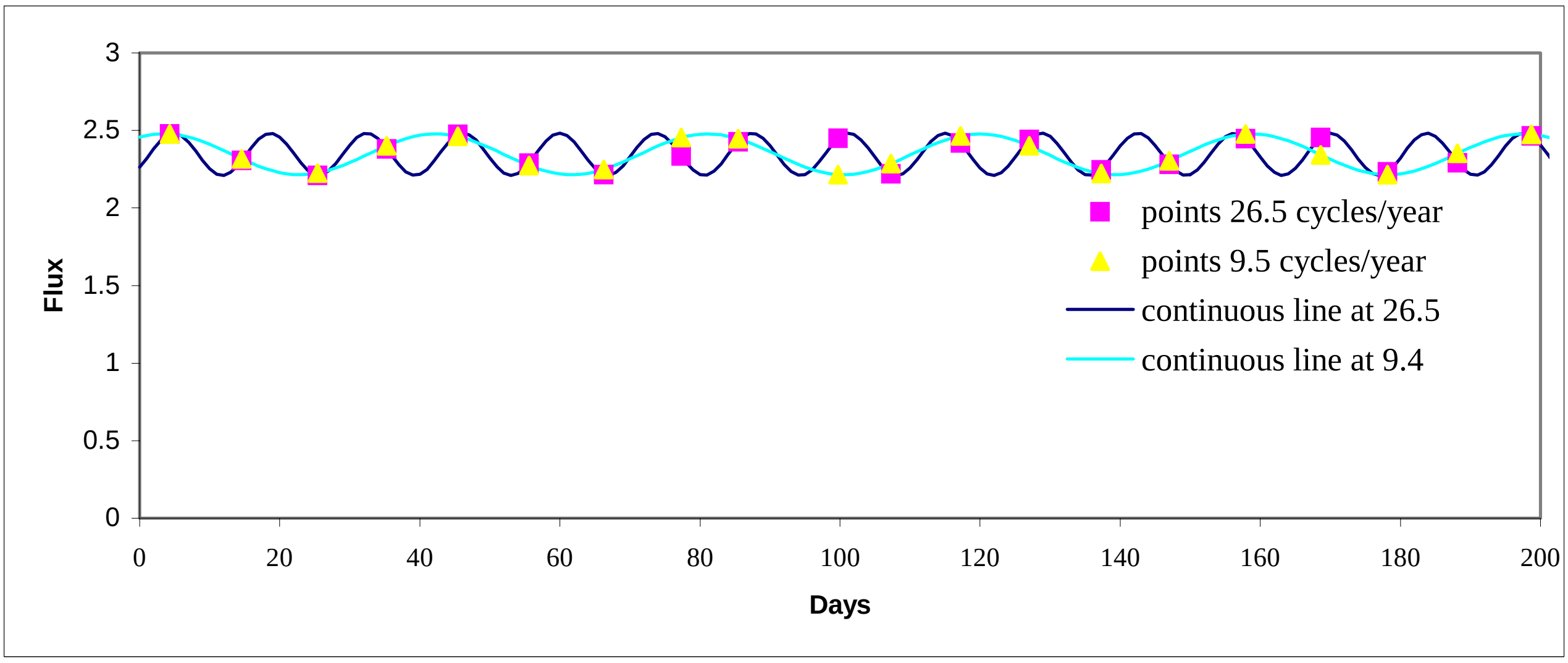

**Figure 4** - *The alias phenomenon showed through the intersection of the two continuous sinusoids of frequency 26.52 and 9.4 cycles/years*

Some words have to be spent for the line at 26.52 cycles/year. This line has been matter of debate in previous analysis of the data [6], [7], especially because it was indicated as a strong modulation candidate. In [7] it was pointed out that this line and that at 9.4 cycles/year constitute an alias pair. It could be instructive to show graphically this fact. In Fig. 3 it is reported an expansion of the time region of the Super-Kamiokande data between 0 and 200 days of data taking: the green points are the experimental flux values, the red and blue points are the values, respectively, of the 9.4 and 26.52 cycles/year modulations inferred from the maximization of the likelihood ratio. It is clear that these two last series of points are very close each other, with a perfect coincidence for some of them.

In Fig. 4 the two 9.4 and 26.52 point series are reported together with the continuous curves to which they belong: the implication of the aliasing is clearly evident in the fact that the two curves intersect each other at most of the sampling points.

It is worth to stress that the alias arises because of the ambiguity due to the undersampling. Indeed from Fig. 4 it stems that if one could be in condition to fill more points between the available sampling points, then the ambiguity would be cancelled.

Before concluding this remark, it has to be added that normally the aliasing to develop requires a series of evenly sampled data, since the time regularity makes it possible to find sinusoids of different frequencies that match the same data points. Actually in such a case the alias is perfect, in the sense that all the points are reproduced exactly by two different modulation signals constituting the alias pair. A prescription often reported to avoid the alias phenomenon is, thus, to introduce some unevenness in the data series; in the present case this does not work because the sampling of the Super-Kamiokande I data maintains a high, even though not perfect, regularity.

## 5. Analysis of the 5 day binned data

The Super-Kamiokande collaboration published the data of the flux organized in 5 day bins with the purpose to check whether there is any recurring pattern in confronting the spectra of the 5 and 10 day binned data. In fact, this kind of comparison can be useful to identify large noise spectral fluctuations, which could appear or disappear depending only upon the binning. In particular, the comparison between the 5 and 10 day data can remove the alias ambiguity between the 9.4 and the 26.52 lines.

The estimated spectrum is shown in Fig. 5, overlapped with the analogous spectrum for the 10 days data. At a first glance it stems that: a) the line at 9.4 (which now is at 9.42 cycles/year) is enhanced with respect to the 10 day case, while the line at 26.52 disappears, thus resolving in favour of the former the alias ambiguity; the fit gives as relative amplitude of the line the value $0.066 \pm 0.14$; b) in the low frequency region the two spectral profiles are very similar, the only notable exception being the line at 9.42; c) together with the 26.52 line, also the other two lines which featured high (greater than 95%) confidence levels in the 10 day spectrum disappear in the 5 day spectrum; d) in the high frequency region there are peaks in the 5 day spectrum either not present or present with lower value in the previous one.

The confidence levels plotted for the more significant peaks are shown in Fig. 6, together with those for the 10 day bin case: the most striking characteristic is the increase of the significance of the peak at 9.42 cycles/year in passing from the previous to the present spectrum. If we focus our attention, as before, to the lines with statistical significance greater than 95%, we note that of the 4 peaks fulfilling this condition in the 10 day bin spectrum, only the 9.42 line survives in the 5 day spectrum. On the other hand, in the 5 day spectrum there is an excess of lines with C.L. above 95% with respect to the 10 day case, mostly located in the high frequency region. A close inspection of these lines shows that those at 31.21, 33.95, 45.78 and 48.35 do not have any counterpart in the 10 day binned spectrum, while those at 39.23 and at 43.66 appears also in the 10 day binned spectrum, even though at the reduced significance of 89% and 88%, respectively.

## 6. Discussion

In principle lines present in both the 5 and 10 day spectra and characterized by high confidence levels should be considered, with a reasonable degree of reliability, significant candidates as possible modulations. In this sense the line at 9.42 is very well suited to be suspected as a not fake signal: indeed it emerges from the analysis of both the datasets, with in particular a C.L. as high as 99.1 % in the more sensitive 5 days binned series.

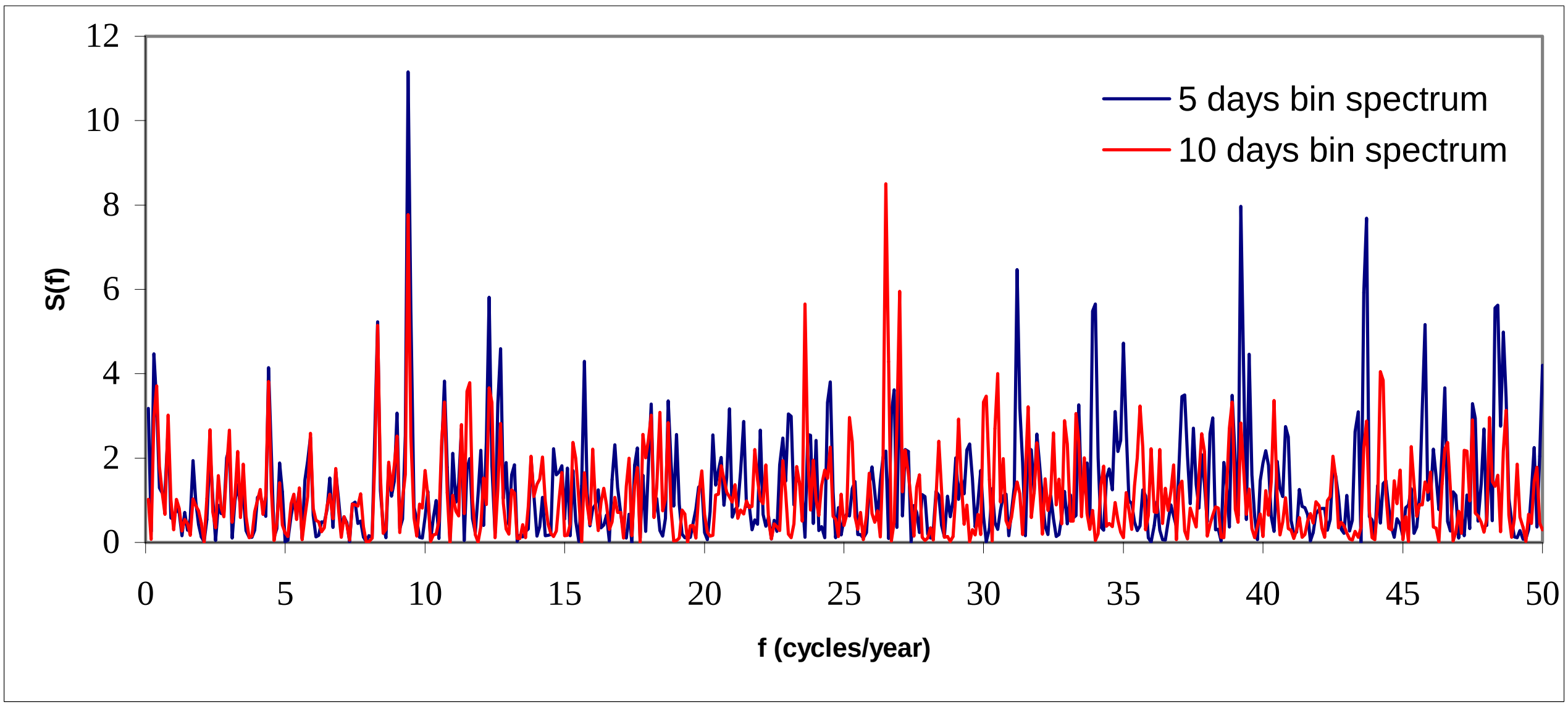

**Figure 5** - *Spectrum of the 5 day binned Super-Kamiokande data series*

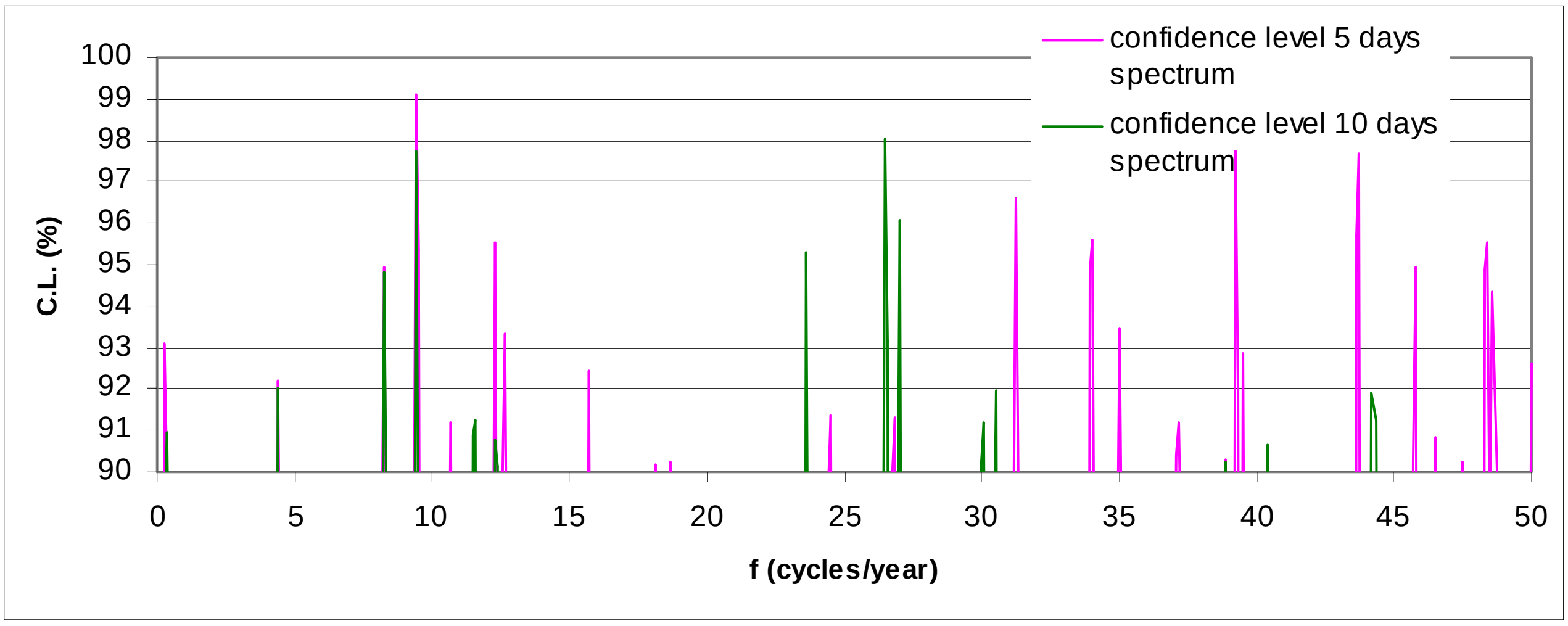

**Figure 6** - *Confidence levels of the more prominent lines in the 5 and 10 day spectra*

For the other lines the diagnostic is much more doubtful. In this respect two features in the mutual comparison of the 5 and 10 day spectra require to be highlighted: a) the presence of a set of recurring lines in the low frequency region. They feature significant confidence levels, even though less than 95%, equal in the two spectra. Obviously, they can well be due to low frequency fluctuations in the data, for which, thus, no change in sensitivity should exist between the two datasets; b) the appearance in the high frequency regime of lines with confidence levels which change radically in passing from one spectrum to the other. In a very conservative interpretation, they can be considered probes of the noisy nature of the spectral estimate, while in a totally different perspective they can be assumed as stemming from the better sensitive to the higher frequency components of the 5 days binned dataset (this interpretation could be in particular reasonable for the two lines at 39.23 and at 43.66, since they appear also in the 10 day case).

Clearly, the conservative approach that tends to attribute the "erratic" high confidence level lines appearing only in the 5 day spectrum to pure noise effects, somehow would also undermine the credibility of the 9.42 component, since would point towards the existence in the spectrum of noisy lines, even though characterized by high confidence levels.

Nevertheless, despite all the caveats induced by the critical inspection of the two spectra and of the associated confidence levels, it is legitimate to point out the special status of the 9.42 cycles/year line, being the only one consistently appearing with confidence level above 95% in the spectral analysis both of the 5 and 10 day binned data series. In a very cautious approach, it can be reliably concluded that the present statistical data scan *fails to reject such a spectral component as a potential modulation candidate*. To ascertain whether this line is a real signal or a large data fluctuation requires other inputs: first of all a solid physical interpretation of its origin, as well as the comparison with the data from a totally independent experiment. In this respect, an interesting test could be provided by the SNO collaboration, by performing a similar scan of their time series data, especially considering that they measure the same $^8$B solar neutrino component as Super-Kamiokande.

**7. Comparison with previous results**

The results reported here can be compared with the recent analysis reported in [8] and [9]. In these papers, however, instead of simply calculating the sinusoidal signal at the mean live time point of each bin, the authors preferred to evaluate an average value over each bin. For comparison purpose the analysis has been repeated with that methodology, thus assuming

$$S_k = \frac{1}{\left(t_{e,k} - t_{s,k}\right)} \int_{t_{s,k}}^{t_{e,k}} A \sin(2\pi f t + \varphi) dt \qquad (18)$$

where $t_{ek}$ and $t_{sk}$ are the set of times provided by the Super-Kamiokande collaboration denoting, respectively, the start and the end of the $k_{th}$ bin. The overlap of the 5 day spectrum evaluated previously with that computed in this alternative way is shown in Fig. 7, from which we can infer that there is no appreciable differences among the two models below about 40 cycles/year, while there is a slight discrepancy in the region above. It is, however, worth point out that the most affected lines by the different modelling are just those at 39.23 and 43.66, mentioned in the previous paragraph: while the former is suppressed, the latter is enhanced by the most refined model.

Numerical comparisons of the major peaks of the 5 day spectrum in [8] with those of the spectrum got here confirm the mutual consistency of the results: positions as well as heights coincide with reasonable accuracy. For example, the peak at 9.42 (located in [8] at 9.43) is found with the analysis of paragraph 5 as having an height of 11.36, which becomes with the analysis based on the model (18) 11.51, that is exactly equal to the value quoted in [8].

The spectral comparison is somehow satisfactory also with the spectrum deduced in [9]: the three major peaks are reported at the frequencies of 9.43, 43.74 and 39.30, which compare well with the same peaks found here. The spectral values are however slightly different: the height of the peak at 9.43 is quoted, for example, equal to 10.88, hence less than what got with the present analysis. The significance of the peaks in the frequency spectrum in [8] and [9] is assessed via Monte Carlo calculations: the authors generated numerous synthetic samples of Super-Kamiokande data, with no modulations embedded, for each of them evaluated the relevant frequency spectrum, and then counted how many times they found the highest spectral line being above the actual value of the 9.42 line.

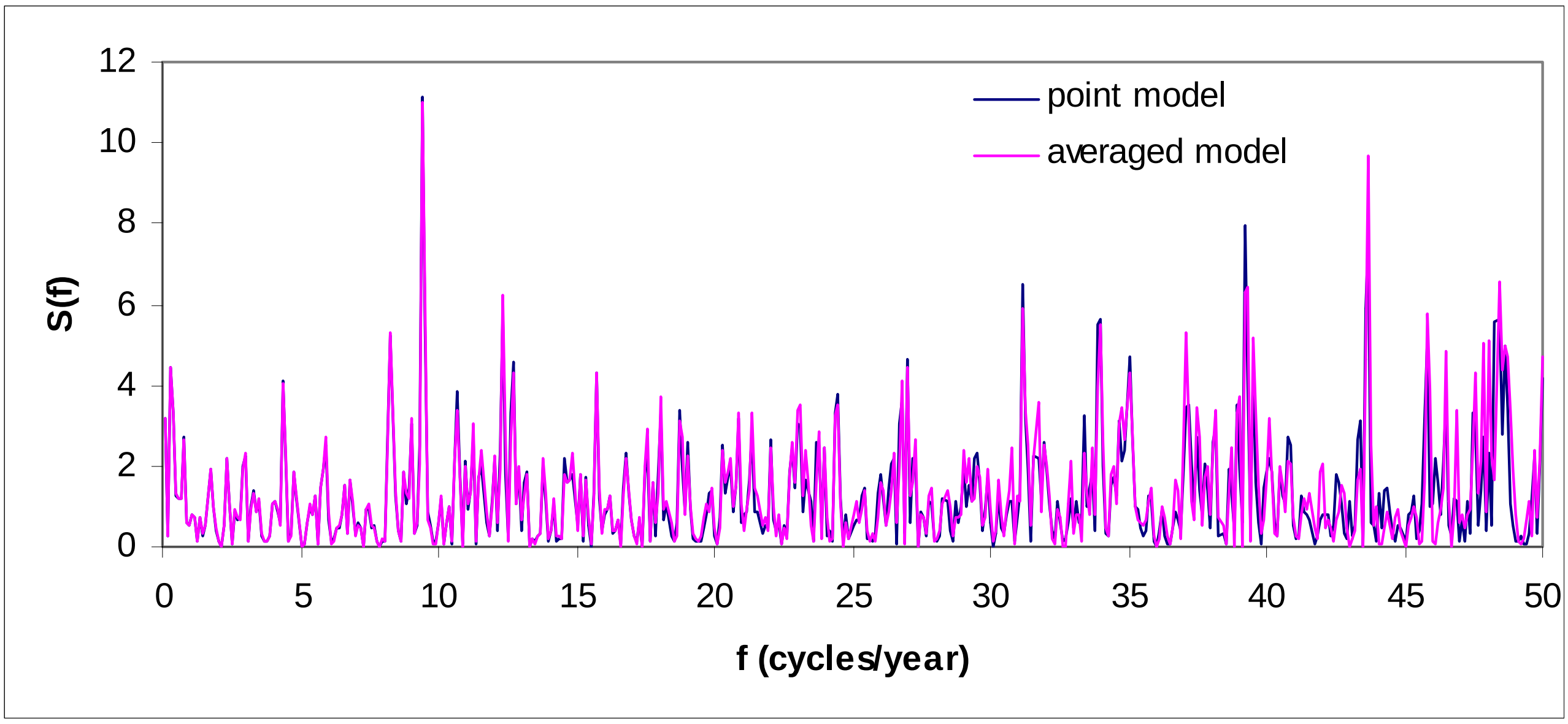

**Figure 7** - *Comparison of the point model with the averaged model for the 5 day bin series*

The value quoted in such a way in [8] for the peak at 9.42 is in the same range of confidence level got here through the application of the expression (16), inferred on the basis of a pure probabilistic reasoning. On the other hand, the same Monte Carlo assessment carried out in [9] leads to quote the same peak as statistically not significant.

Hence in summary, the methodology proposed here is in a good spectral agreement with both the previous analysis on the Super-Kamiokande I data, while there is agreement for the significance assessment only with the analysis in [8]. The understanding of the origin of the discrepancy with respect to the analysis in [9] cannot be performed in the framework of the present work, based explicitly on an analytical approach which avoids the Monte Carlo calculation.

## 8. Relationship with the Lomb Scargle method

The methodology developed here can be put in direct comparison with the classical Lomb Scargle method. To do so, it is preliminary required to modify the Lomb Scargle periodogram formulation in order to take into account the error information individually available for each point of the data series. The explicit formulation of this modified form of the periodogram, as derived in the appendix, is

$$\frac{1}{2}\frac{\left(\sum_{k=1}^{N}\frac{(x_k-F)\cos[\omega(t_k-\tau)]}{\sigma_k^2}\right)^2}{\sum_{k=1}^{N}\frac{\cos^2[\omega(t_k-\tau)]}{\sigma_k^2}}+\frac{1}{2}\frac{\left(\sum_{k=1}^{N}\frac{(x_k-F)\sin[\omega(t_k-\tau)]}{\sigma_k^2}\right)^2}{\sum_{k=1}^{N}\frac{\sin^2[\omega(t_k-\tau)]}{\sigma_k^2}} \qquad (19)$$

where $\tau$ is given by

$$\frac{\sum_{k=1}^{N}\frac{\sin 2\omega t_k}{\sigma_k^2}}{\sum_{k=1}^{N}\frac{\cos 2\omega t_k}{\sigma_k^2}}=\tan 2\omega\tau \qquad (20)$$

and as usual $\omega = 2\pi f$ ; the (19) reduces to the standard Lomb Scargle periodogram definition (see for example [11]) when the $\sigma_k$ are all equal.

Furthermore, in the appendix along the derivation of this formula it is demonstrated also that the periodogram so obtained is a numerical approximation of the log-likelihood ratio (13) when evaluated at each frequency by inserting the corresponding amplitude and phase parameters which maximize the likelihood function. Since this specific evaluation of the log-likelihod ratio (13) is just the way the spectrum of the data series is defined in the framework of the likelihood method, it turns out hence that the Lomb Scargle modified periodogram is an approximation of the likelihood spectrum itself. Actually, it is a very good approximation: indeed, as an example, the Lomb Scargle modified periodogram of the 5 day binned SK data has been computed and compared with the corresponding likelihood spectrum in figure 5, and they resulted to be practically indistinguishable. Moreover, it has to be pointed that the (19) features the same distribution properties of the standard Lomb-Scargle periodogram (also this demonstration is in the appendix). This means that in the case of null hypothesis (i.e. no modulation embedded in the series) the (19) is distributed simply as $e^{-z}$ [2]. This circumstance allows to use the false alarm probability formula introduced in [2] (see also [11]) to perform (approximate) peaks significance assessments. The false alarm formula is

$$P(>z) = 1-(1-e^{-z})^M \qquad (21)$$

where M is the number of independent scanned frequencies; according to the prescription reported in [11], M is equal to the number N of data points if the range scanned is up to the Nyquist frequency, while it increases proportionally as the scanning range increases beyond that limit.

In our case, since the Nyquist frequency is 36 cycles/year and we scan up to 50 cycles/year, and N is equal to 358 data points, then M is equal to 497. Inserting this value in (21), as well as the z maximum value of the Lomb-Scargle modified periodogram at the peak frequency of 9.42 cycles/year, which is 11.36, we get as false alarm probability the value of 0.6%, or equivalently a significance of the peak of 99.4%, in good agreement with the value reported in paragraph 6, obtained on the basis of the likelihood method.

**9. Some consistency checks on the 5 day bin dataset**

It could be useful to perform some overall checks in order to get a direct feeling of the consistency of the method and of the results obtained.

A first check can be done through the study of the annual modulation expected from the Earth’s orbit eccentricity, instead of removing it from the data. For example, one may use the data not corrected with the compensation factors provided by Super-Kamiokande to assess the effectiveness of the likelihood method (or equivalently of the Lomb Scargle modified periodogram) with respect to the standard Lomb Scargle periodogram in capturing the 1 cycles/year line. In Fig. 8 it is shown the overlap in the frequency region from 0 to 5 cycles/year of the two likelihood spectra related to the corrected and not corrected data. It is clearly visible the 1 cycles/year line in the spectrum of the not corrected data, otherwise absent in the other spectrum. The corresponding C.L.is 93.5%. Furthermore, in Fig. 9 it is shown the overlap, in the same frequency range as before, of the Lomb Scargle modified periodogram (which is the same to say the likelihood spectrum) with the standard Lomb Scargle periodogram: the former clearly captures more efficiently the annual modulation line, thus proving that properly taking into account the errors is important in this context.

One, however, may wonder whether the line is captured properly. The answer is in Fig. 10: the two plots represent, respectively, the annual modulation as provided by the Super-Kamiokande collaboration on the basis of calendar dates of the data taking and the annual modulation inferred by the best fit amplitude and phase parameters get through the likelihood maximization. The two plots show a reasonable agreement.

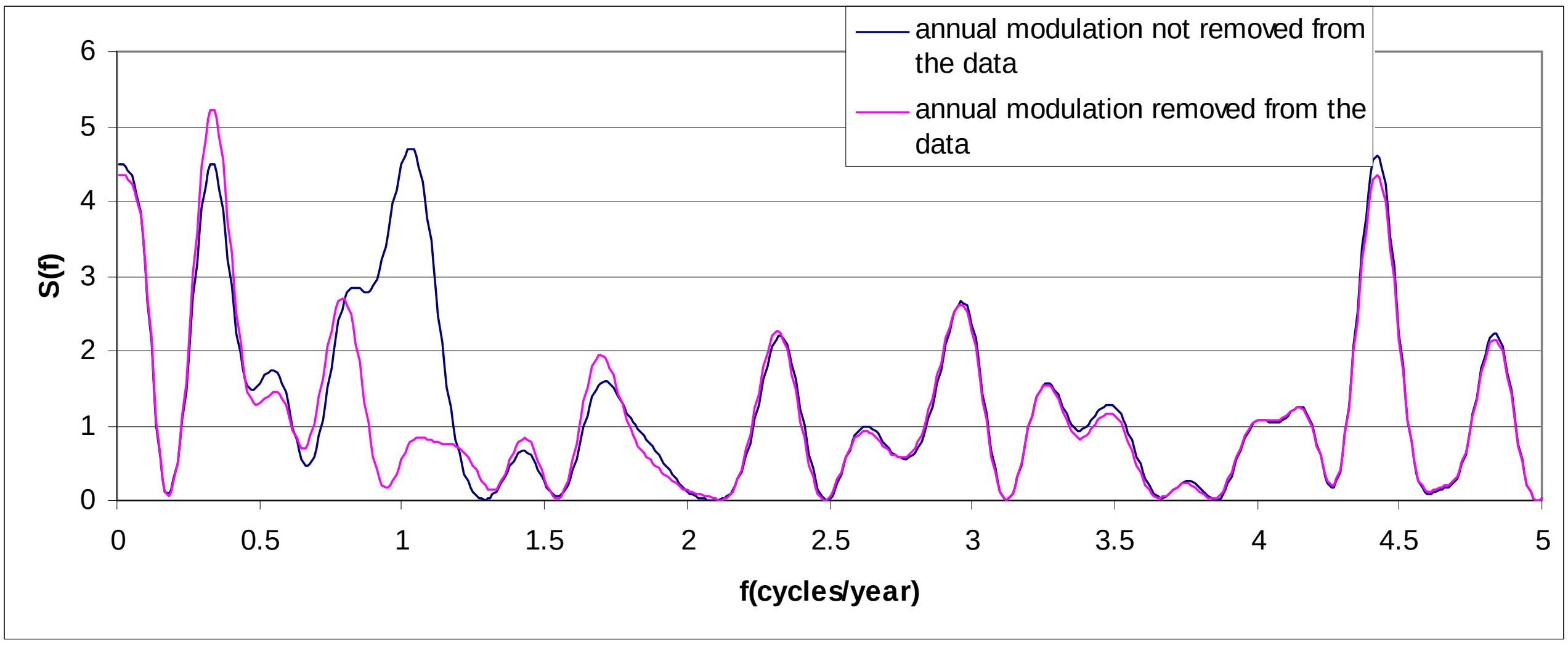

**Figure 8** - *Comparison of the spectra with annual modulation removed and not removed*

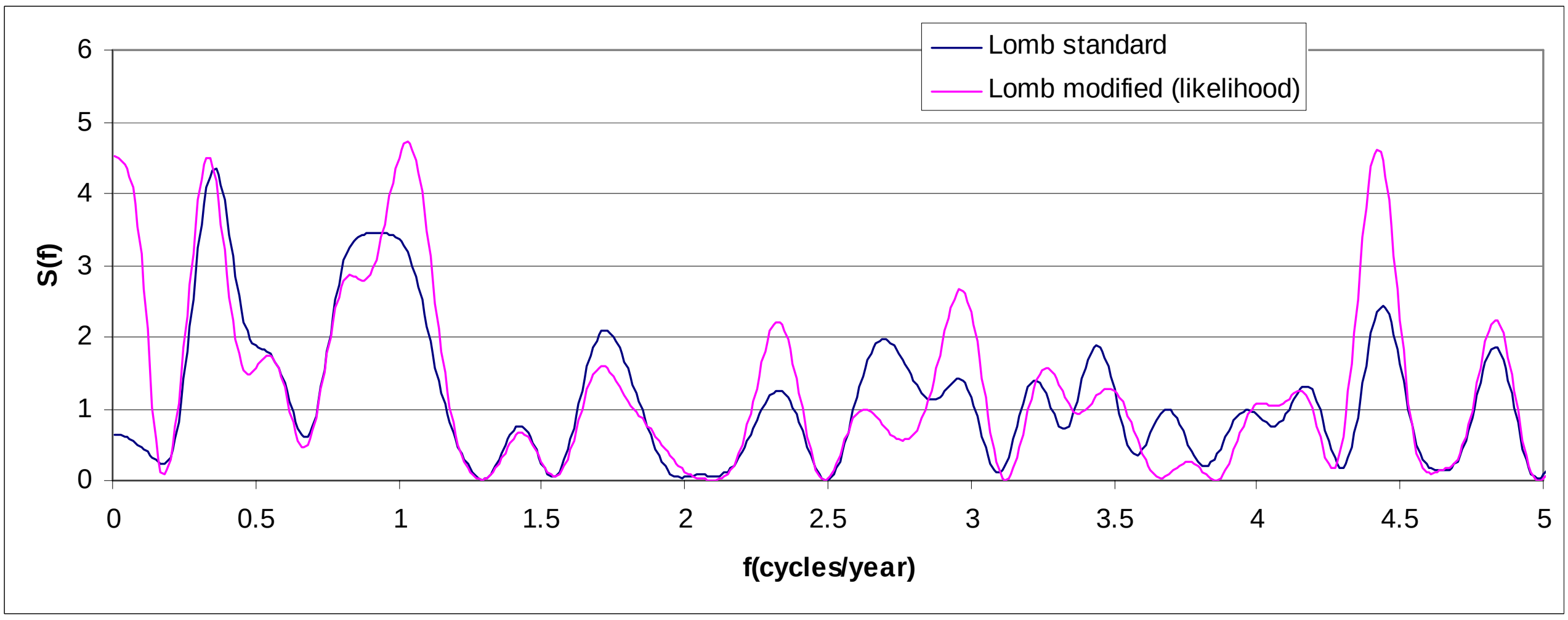

**Figure 9** – *Annual modulation not removed: comparison of the Lomb standard and Lomb modified (likelihood) spectra*

The relative amplitude of the modulation is, in particular, evaluated equal to 0.042 ± 0.014, to be compared with the nominal value of 0.034. While these results seem somehow reassuring, one may asks why the annual line is quoted at a confidence level of 93.5%, which could appear low considering that we are dealing with a signal surely present in the data. The reason is connected to the usual signal to noise ratio effect: the amplitude of the annual modulation compared to the noise affecting the data cannot produce a more outstanding peak in the frequency spectrum, featuring an higher confidence level.

So, in summary, the algorithm is able to detect reasonably the features of a modulation actually embedded in the series, even if overwhelmed by noise. However, given the noise level, such a line could not be claimed as a true signal, without actually knowing that it must be there.

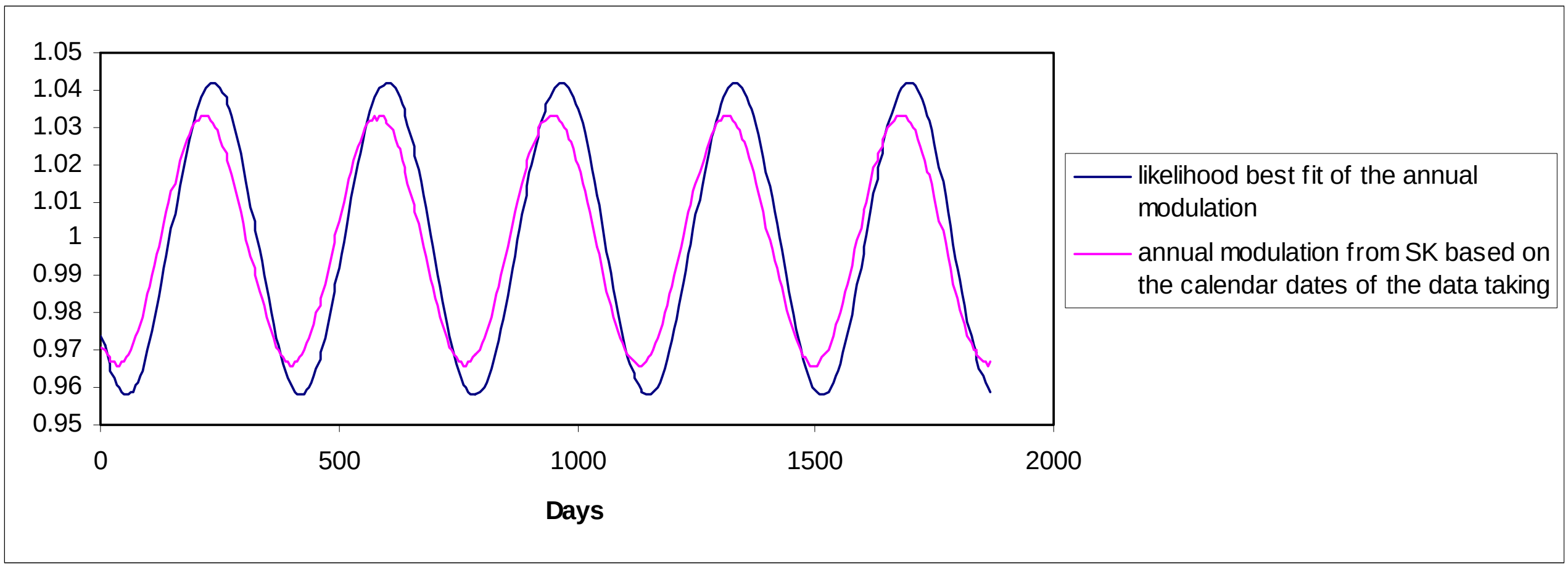

**Figure 10** – *Estimated vs. true annual modulation*

Other possible cross checks concern the presence of alias frequencies in the spectrum. In the case of data totally unevenly sampled the alias phenomenon should not occur (see for example [11]). However, the sampling time sequence of the Super-Kamiokande data maintains a certain degree of regularity that was already noted in the 10 day bin data, just through the associate alias phenomenon.

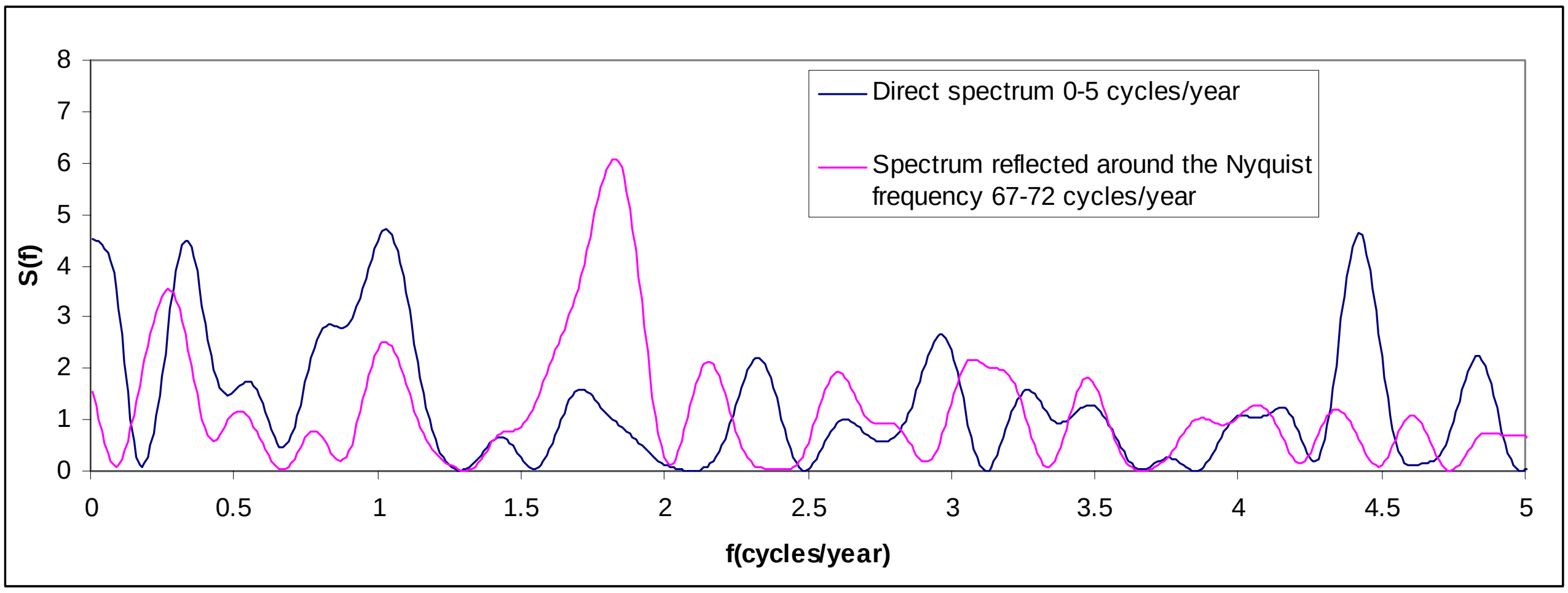

**Figure 11** – *Spectrum of the data not corrected for the annual modulation: 0-5 frequency region overlapped to the 67-72 image frequency region. The annual line and its possible alias are clearly visible*

It is worth to remind that the alias effect develops as an interaction of the sampling frequency with the frequencies of the modulation signals present in the data series, producing ghost frequencies of the real ones, which are specularly located with respect to the Nyquist frequency. Considering that the nominal Nyquist frequency is equal to about 36 cycles/year, the alias of the 1 cycles/year line should be located at the frequency of about 36+(36-1)=71 cycles/year. In order to make it easy the identification of the alias frequency, in the relevant figures the curves are plotted overlapping a frequency region below the Nyquist frequency with the respective, reversed, image region above the Nyquist limit (so, for example, the region from 0 to 5 cycles/year is overlapped to the region from 67 to 72 cycles/year, plotted in reverse order so that 0 coincide with 72 and 5 with 67): in this way the frequency of a real signal and its alias will coincide on the abscissa.

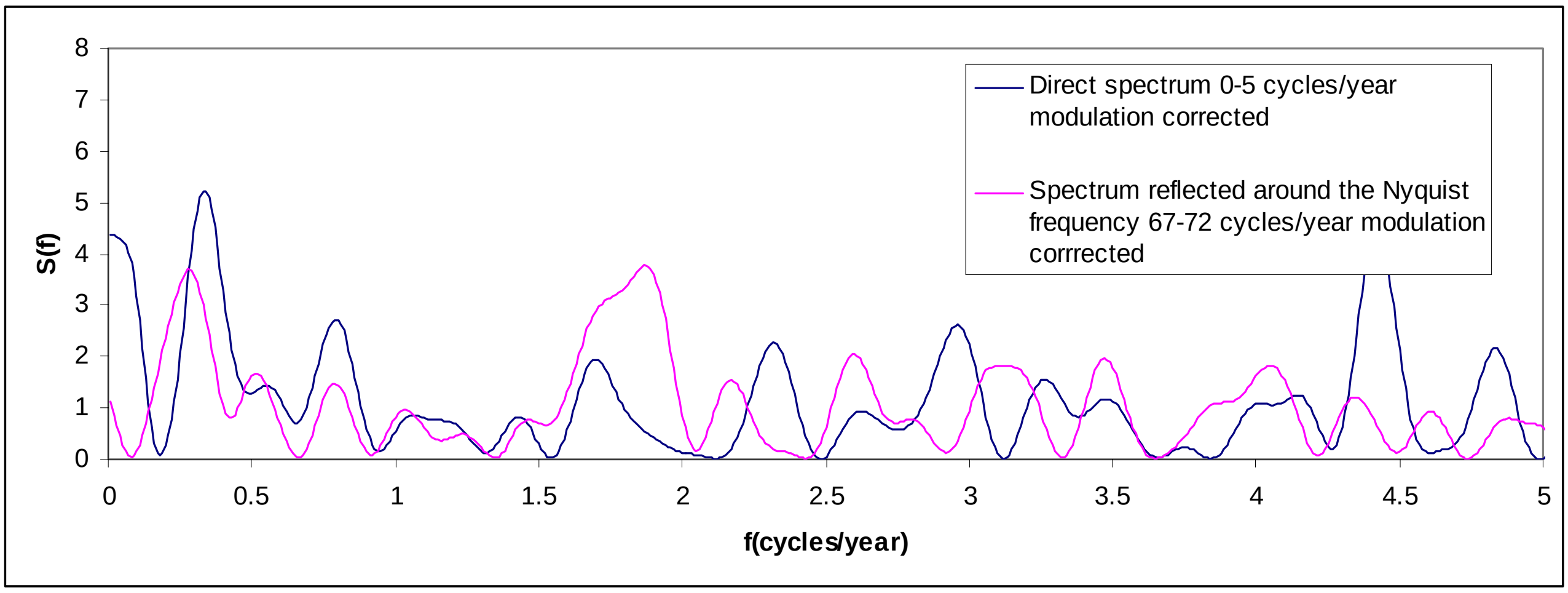

**Figure 12** - *Spectrum of the data corrected for the annual modulation: 0-5 frequency region overlapped to the 67-72 image frequency region. Both the annual line and the corresponding potential alias disappear*

In Fig.11 this kind of graphical representation is reported still focusing on the frequency region from 0 to 5 cycles/year to highlight again the behaviour of the 1 cycles/year modulation. The annual line in the blue plot is that already studied above. In correspondence to it there is in the other plot (i.e. that related to the image frequency region) a less high line that could be the alias. Is it truly the alias? An easy check can be done examining in the same frequency window the spectrum obtained from the data corrected for the annual modulation effect trough the SK corrective factors. The 1 cycles/year line is already known to disappear; if the corresponding frequency line is the alias, it should disappear as well. Fig. 12 gives the answer: both the lines in the two plots vanish, so it can be concluded that the regularity of the sampling actually gives rises to the alias of a frequency surely present in the data.

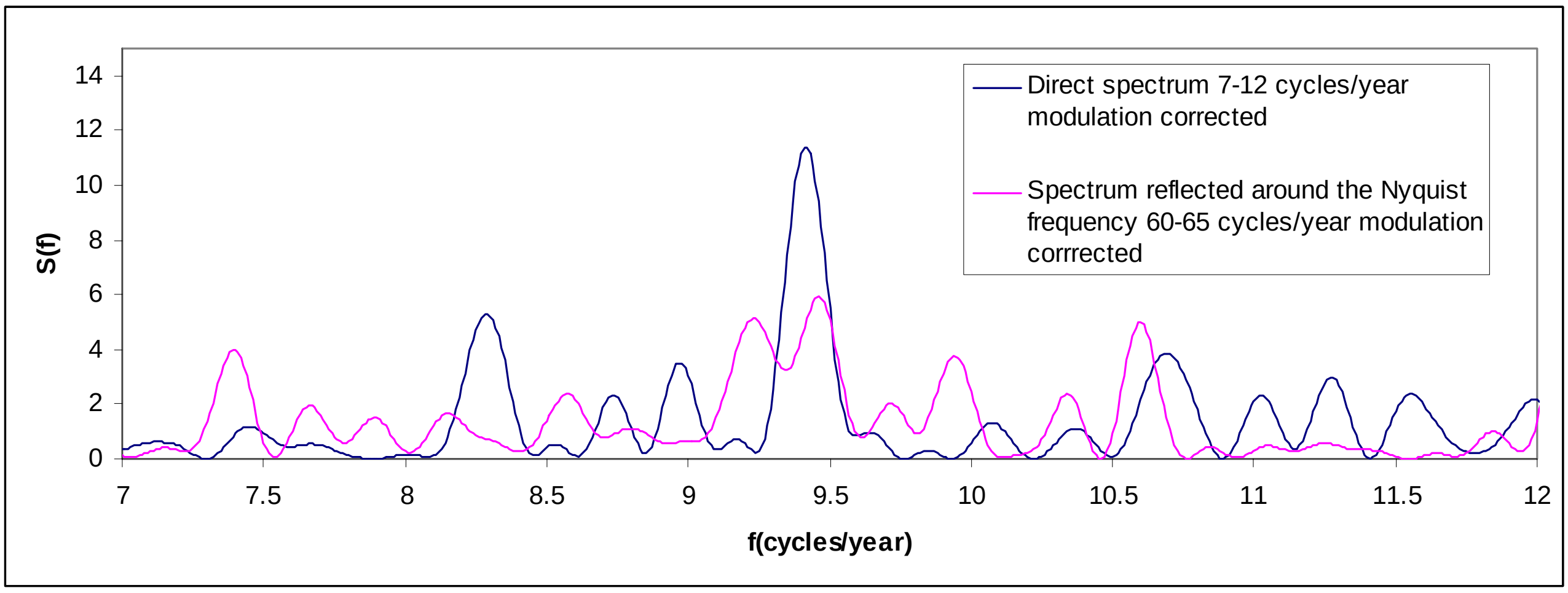

**Figure 13** - *7-12 frequency region overlapped to the 60-65 image frequency region. In correspondence of the 9.42 line there is a peak in the image frequency plot.*

In doing this check, it has also been verified that the effective Nyquist frequency, that makes the two lines coincide exactly, is 39.95 cycles/year, thus close to the nominal value.

It can now be checked what happens to the 9.42 line; for this purpose the same graph is now plotted in the region from 7 to 12 cycles/year and reported in Fig 13. In correspondence of the 9.42 line there is actually a peak in the plot related to the image frequency region, that could well be the alias, thus essentially reproducing the same situation discussed at length in the framework of the 10 day bin data. The small relative shift between the two peaks could be likely due to the imperfect sampling.

In conclusion, these methodological checks provide some evidence of the overall consistency both of the method and of the results. Still, a conservative “sceptical” observer would continue not to do any claim about the reality of the suspect 9.42 line, but he/she would simply take note that also these verifications fail to falsify the suspect line: indeed, if after the notice of the existence of the alias of the annual modulation, no potential alias line would have appeared for the 9.42 frequency, it could have been obviously concluded that such a line were almost surely a data fluctuation. On the other hand, given the obtained result, the “falsification” of this line will have to wait for additional inputs.

## 10. Conclusions

In this work, besides reviewing the overall properties of the likelihood methodology applied to the scan of time series data, it has been in particular evaluated a “false alarm” formula that, in total analogy with the classical Lomb-Scargle method, allows to compute analytically without invoking Monte Carlo calculations the significance of the modulation signatures identified via the likelihood approach.

After examining the features of the likelihood method, it has also been shown its close relationship with the Lomb Scargle methodology; in particular it has been derived a modification of the standard Lomb Scargle periodogram which takes properly into account the individual errors of the terms of the data series, and it has been demonstrated that it represents a very close approximation of the likelihood spectrum.

The application, as example, of the method to the 5 and 10 day binned Super-Kamiokande data produces a spectral distribution whose most outstanding characteristic is a line at a frequency of 9.42 cycles/year, with confidence level of about 99.1% as stemming from the analysis of the 5 day bin data set. However, the proper interpretation of this result must be stressed clearly: *in the Super-Kamiokande data there is a feature which statistics alone, as applied here, cannot disprove as a potential modulation*. To go forward and disprove or prove this line, therefore, is not a matter only of statistics, but would require additional experimental inputs, as for example a similar analysis coming from another real time solar neutrino experiment, and it would not be surprising at all if, eventually, it should prove to be only a sort of large fluctuation in the data.

It is worth point out that also the Lomb-Scargle modified periodogram peak significance assessment, performed in analogy with the standard Lomb Scargle method, results in a significance of the 9.42 peak of more than 99% (actually 99.4%), thus representing a consistent cross check of the overall methodology depicted here.

In the pure statistical framework, it remains open the issue of the different significance for the same 9.42 cycles/year spectral line evaluated here with the analytical approach and in [9] through the Monte Carlo calculation, while the analysis in [8], based on Monte Carlo as well, agrees with the present findings.

Finally, a further check of the consistency of the calculations has been attempted exploiting both the presence of the annual modulation frequency in the spectrum and the appearance of alias effects, getting overall coherent results.

**Acknowledgments**

The author would like to thank Raju Raghavan for many useful discussions, David J. Thomson for important remarks, Luciano Pandola for having red a preliminary version of this manuscript, Eligio Lisi for suggesting the consistency cross checks reported in paragraph 9, and the Super-Kamiokande Collaboration for making the data publicly available.

APPENDIX – ANALYTICAL COMPARISON BETWEEN THE LIKELIHOOD METHOD AND A MODIFIED FORM OF THE LOMB-SCARGLE METHOD

Purpose of this appendix is to demonstrate the relationship between the likelihood based methodology and the Lomb-Scargle method modified to take into account the experimental errors on each data point (whereas the standard Lomb-Scargle periodogram does not exploit such an information).

To this end, let's follow for simplicity the notation of the original Lomb paper [1], thus assuming oscillations around zero written as

$$A\cos\omega t + B\sin\omega t \tag{A1}$$

hence if the series $x_k^{'}$ features a non zero average $F$ , then it is simply replaced by

$$x_k = x_k^{'} - F \tag{A2}$$

The likelihood is thus

$$L = e^{-\frac{1}{2}\left(\sum_{k=1}^{N}\frac{\left[x_k - \left(A\cos\omega t_k + B\sin\omega t_k\right)\right]^2}{\sigma_k^2}\right)} \tag{A3}$$

which is maximized when the term

$$E = \sum_{k=1}^{N}\frac{\left[x_k - \left(A\cos\omega t_k + B\sin\omega t_k\right)\right]^2}{\sigma_k^2} \tag{A4}$$

gets the minimum. The minimization of (A4) is the immediate extension of the least square un-weighted fit to sinusoids adopted by Lomb to derive the periodogram in the cases in which the $\sigma_k$ are not available. Furthermore, writing the likelihood ratio as

$$L = \frac{e^{-\frac{1}{2}\left(\sum_{k=1}^{N}\frac{\left[x_k - \left(A\cos\omega t_k + B\sin\omega t_k\right)\right]^2}{\sigma_k^2}\right)}}{e^{-\frac{1}{2}\sum_{k=1}^{N}\frac{x_k^2}{\sigma_k^2}}} \tag{A5}$$

and following the spectrum definition introduced in paragraph 3, we get

$$S = \frac{1}{2}\sum_{k=1}^{N}\frac{x_k^2}{\sigma_k^2} - \frac{1}{2}\sum_{k=1}^{N}\frac{\left[x_k - \left(A\cos\omega t_k + B\sin\omega t_k\right)\right]^2}{\sigma_k^2} \tag{A6}$$

which represents the natural extension, that accounts for the presence of the $\sigma_k$ factors, of the original Lomb's spectrum definition (definition stemming from the previous work of Barning [10]).

In order to study the statistical properties of (A6) we can follow the same minimization procedure used by Lomb, generalized to the current case in which the $\sigma_k$ are present.

First we proceed to minimize (A4) by performing the partial derivatives

$$\frac{\partial E}{\partial A} = \sum_{k=1}^{N} \frac{2}{\sigma_k^2} \left[ x_k - \left( A \cos \omega t_k + B \sin \omega t_k \right) \right] \left( - \cos \omega t_k \right) \tag{A7}$$

$$\frac{\partial E}{\partial B} = \sum_{k=1}^{N} \frac{2}{\sigma_k^2} \left[ x_k - \left( A \cos \omega t_k + B \sin \omega t_k \right) \right] \left( - \sin \omega t_k \right) \tag{A8}$$

and equating both equations to zero we get

$$A \sum_{k=1}^{N} \frac{\cos^2 \omega t_k}{\sigma_k^2} + B \sum_{k=1}^{N} \frac{\sin \omega t_k \cos \omega t_k}{\sigma_k^2} = \sum_{k=1}^{N} \frac{x_k}{\sigma_k^2} \cos \omega t_k \tag{A9}$$

$$A \sum_{k=1}^{N} \frac{\sin \omega t_k \cos \omega t_k}{\sigma_k^2} + B \sum_{k=1}^{N} \frac{\sin^2 \omega t_k}{\sigma_k^2} = \sum_{k=1}^{N} \frac{x_k}{\sigma_k^2} \sin \omega t_k \tag{A10}$$

which can be written in matrix notation

$$\begin{pmatrix} \sum_{k=1}^{N} \frac{\cos^2 \omega t_k}{\sigma_k^2} & \sum_{k=1}^{N} \frac{\sin \omega t_k \cos \omega t_k}{\sigma_k^2} \\ \sum_{k=1}^{N} \frac{\sin \omega t_k \cos \omega t_k}{\sigma_k^2} & \sum_{k=1}^{N} \frac{\sin^2 \omega t_k}{\sigma_k^2} \end{pmatrix} \begin{pmatrix} A \\ B \end{pmatrix} = \begin{pmatrix} \sum_{k=1}^{N} \frac{x_k}{\sigma_k^2} \cos \omega t_k \\ \sum_{k=1}^{N} \frac{x_k}{\sigma_k^2} \sin \omega t_k \end{pmatrix} \tag{A11}$$

By defining

$$cc = \sum_{k=1}^{N} \frac{\cos^2 \omega t_k}{\sigma_k^2} \quad cs = \sum_{k=1}^{N} \frac{\sin \omega t_k \cos \omega t_k}{\sigma_k^2} \quad ss = \sum_{k=1}^{N} \frac{\sin^2 \omega t_k}{\sigma_k^2} \tag{A12}$$

$$\Delta = \begin{pmatrix} cc & cs \\ cs & ss \end{pmatrix} \tag{A13}$$

we get finally

$$\begin{pmatrix} A \\ B \end{pmatrix} = \Delta^{-1} \begin{pmatrix} \sum_{k=1}^{N} \frac{x_k}{\sigma_k^2} \cos \omega t_k \\ \sum_{k=1}^{N} \frac{x_k}{\sigma_k^2} \sin \omega t_k \end{pmatrix} \tag{A14}$$

being

$$\Delta^{-1} = \begin{pmatrix} \frac{ss}{D} & -\frac{cs}{D} \\ -\frac{cs}{D} & \frac{ss}{D} \end{pmatrix} \qquad D = cc \cdot ss - cs^2 \qquad \text{(A15).}$$

Before inserting the A and B values so computed in (A6), we can manipulate (A6) itself as follows

$$\frac{1}{2}\sum_{k=1}^{N}\frac{x_k^2}{\sigma_k^2} - \frac{1}{2}\sum_{k=1}^{N}\frac{x_k^2}{\sigma_k^2} - \frac{1}{2}\sum_{k=1}^{N}\frac{(A\cos\omega t_k + B\sin\omega t_k)^2}{\sigma_k^2} + \sum_{k=1}^{N}\frac{x_k(A\cos\omega t_k + B\sin\omega t_k)}{\sigma_k^2} \qquad \text{(A16)}$$

which becomes

$$\sum_{k=1}^{N}\frac{x_k(A\cos\omega t_k + B\sin\omega t_k)}{\sigma_k^2} - \frac{1}{2}\sum_{k=1}^{N}\frac{(A\cos\omega t_k + B\sin\omega t_k)^2}{\sigma_k^2} \qquad \text{(A17)}$$

and

$$\sum_{k=1}^{N}\frac{[2x_k - (A\cos\omega t_k + B\sin\omega t_k)](A\cos\omega t_k + B\sin\omega t_k)}{2\sigma_k^2} \qquad \text{(A18).}$$

By writing the last expression as

$$\sum_{k=1}^{N}\frac{\{x_k + [x_k - (A\cos\omega t_k + B\sin\omega t_k)]\}(A\cos\omega t_k + B\sin\omega t_k)}{2\sigma_k^2} \qquad \text{(A19)}$$

it is easily recognized that the term in brackets $[x_k - (A\cos\omega t_k + B\sin\omega t_k)]$, being the residual of the fit, is negligible with respect to $x_k$. The (A19) thus becomes approximately equal to

$$\sum_{k=1}^{N}\frac{x_k(A\cos\omega t_k + B\sin\omega t_k)}{2\sigma_k^2} \qquad \text{(A20)}$$

that in matrix form can be written

$$\frac{1}{2}\begin{pmatrix} \sum_{k=1}^{N}\frac{x_k\cos\omega t_k}{\sigma_k^2} & \sum_{k=1}^{N}\frac{x_k\sin\omega t_k}{\sigma_k^2} \end{pmatrix}\begin{pmatrix} A \\ B \end{pmatrix} \qquad \text{(A21)}$$

or, according to (A14) and (A15)

$$\frac{1}{2}\begin{pmatrix} \sum_{k=1}^{N}\frac{x_k\cos\omega t_k}{\sigma_k^2} & \sum_{k=1}^{N}\frac{x_k\sin\omega t_k}{\sigma_k^2} \end{pmatrix}\begin{pmatrix} \frac{ss}{D} & -\frac{cs}{D} \\ -\frac{cs}{D} & \frac{cc}{D} \end{pmatrix}\begin{pmatrix} \sum_{k=1}^{N}\frac{x_k\cos\omega t_k}{\sigma_k^2} \\ \sum_{k=1}^{N}\frac{x_k\sin\omega t_k}{\sigma_k^2} \end{pmatrix} \qquad \text{(A22).}$$

Following Lomb, the 2x2 matrix in (A22) can be put in diagonal form making *cs=0.* As shown in [1], this is obtained inserting a time shift such that (A1) becomes

$$A\cos\omega(t-\tau)+B\sin\omega(t-\tau) \qquad \text{(A23)}$$

where $\tau$ is deduced from the equation

$$\frac{\sum_{k=1}^{N}\frac{\sin 2\omega t_k}{\sigma_k^2}}{\sum_{k=1}^{N}\frac{\cos 2\omega t_k}{\sigma_k^2}}=\tan 2\omega\tau \qquad \text{(A24).}$$

With this choice the (A22) becomes

$$\frac{1}{2}\begin{pmatrix}\sum_{k=1}^{N}\frac{x_k\cos\omega t_k}{\sigma_k^2} & \sum_{k=1}^{N}\frac{x_k\sin\omega t_k}{\sigma_k^2}\end{pmatrix}\begin{pmatrix}\frac{1}{cc} & 0\\ 0 & \frac{1}{ss}\end{pmatrix}\begin{pmatrix}\sum_{k=1}^{N}\frac{x_k\cos\omega t_k}{\sigma_k^2}\\ \sum_{k=1}^{N}\frac{x_k\sin\omega t_k}{\sigma_k^2}\end{pmatrix} \qquad \text{(A25)}$$

which can be manipulated to obtain

$$\frac{1}{2}\begin{pmatrix}\frac{1}{cc}\sum_{k=1}^{N}\frac{x_k\cos\omega t_k}{\sigma_k^2} & \frac{1}{ss}\sum_{k=1}^{N}\frac{x_k\sin\omega t_k}{\sigma_k^2}\end{pmatrix}\begin{pmatrix}\sum_{k=1}^{N}\frac{x_k\cos\omega t_k}{\sigma_k^2}\\ \sum_{k=1}^{N}\frac{x_k\sin\omega t_k}{\sigma_k^2}\end{pmatrix} \qquad \text{(A26)}$$

$$\frac{1}{2}\left[\frac{1}{cc}\left(\sum_{k=1}^{N}\frac{x_k\cos\omega t_k}{\sigma_k^2}\right)^2+\frac{1}{ss}\left(\sum_{k=1}^{N}\frac{x_k\sin\omega t_k}{\sigma_k^2}\right)^2\right] \qquad \text{(A27)}$$

and finally, remembering (A12)

$$\frac{1}{2}\frac{\left(\sum_{k=1}^{N}\frac{x_k\cos\omega t_k}{\sigma_k^2}\right)^2}{\sum_{k=1}^{N}\frac{\cos^2\omega t_k}{\sigma_k^2}}+\frac{1}{2}\frac{\left(\sum_{k=1}^{N}\frac{x_k\sin\omega t_k}{\sigma_k^2}\right)^2}{\sum_{k=1}^{N}\frac{\sin^2\omega t_k}{\sigma_k^2}} \qquad \text{(A28)}$$

where, because of the (A23), the $t_k$ actually stand for ($t_k$ - $\tau$) (and for a not zero averaged series $x_k$ is replaced by $x_k = x_k^{'} - F$ ).

The (A28) is the generalization of the Lomb-Scargle periodogram which takes into account the errors associated with each data point. It reduces to the standard Lomb-Scargle periodogram if all the $\sigma_k$ are equal, and features the same distribution properties. Indeed, following [2], in the case of purely noisy series (i.e. no modulation embedded) (A28) can be considered the sum of the squares of two normally distributed zero mean random variables, i.e.

$$\left( \frac{\sum_{k=1}^{N} \frac{x_k \cos \omega t_k}{\sigma_k^2}}{\sqrt{2\sum_{k=1}^{N} \frac{\cos^2 \omega t_k}{\sigma_k^2}}} \right)^2 + \left( \frac{\sum_{k=1}^{N} \frac{x_k \sin \omega t_k}{\sigma_k^2}}{\sqrt{2\sum_{k=1}^{N} \frac{\sin^2 \omega t_k}{\sigma_k^2}}} \right)^2 \qquad \text{(A29).}$$

The terms in the brackets are linear combinations of the variables $x_k$: since they are zero mean and normally distributed variables, this ensures that the same property is featured also by their combinations.

The term within the first bracket features a variance which is equal to

$$\frac{\sum_k \sum_j \frac{\langle x_k x_j \rangle \cos \omega t_k \cos \omega t_j}{\sigma_k^2 \sigma_j^2}}{2\sum_{k=1}^{N} \frac{\cos^2 \omega t_k}{\sigma_k^2}} = \frac{1}{2} \qquad \text{(A30)}$$

(this result is obtained considering that $\langle x_k x_j \rangle = 0$ if $k \neq j$ and $\langle x_k x_j \rangle = \sigma_k^2$ if $k = j$).

Similarly, also the variance of the variable in the second bracket of (A26) is equal to ½. It can now be used the result that the distribution of the sum of the squares of two gaussian random variables with zero mean and equal variance $\sigma$ is [2]

$$\frac{1}{2\sigma^2} e^{-\frac{z}{2\sigma^2}} \qquad \text{(A31)}$$

and, since in the present case $\sigma^2$ is ½, we get the result that the distribution of the modified Lomb-Scargle periodogram is simply $e^{-z}$, like that of the standard Lomb-Scargle periodogram [2] (it is worth to repeat this result is valid if there is no modulation in the time series).

From the overall deduction presented in this appendix it stems clearly the close relationship between the spectrum obtained via the Lomb-Scargle prescription (modified to take into account the experimental errors) given by the (A28), and the spectrum inferred by the application of the likelihood ratio ((A6) or equivalently the (13) of the paragraph 3): *the former is a numerical approximation of the latter*. This approximation practically proves to be very effective, both in terms of accuracy and of speed of calculation.